\documentclass[12pt]{article}

\usepackage{amssymb}
\usepackage{amsmath}
\usepackage{amsfonts}

\newcommand{\be}{\begin{equation}}
\newcommand{\bea}{\begin{eqnarray}}
\newcommand{\eea}{\end{eqnarray}}

\newcommand{\ed}{\end{document}}

\begin{document}

\title{Is Pseudo-Hermitian Quantum Mechanics an Indefinite-Metric
Quantum Theory?}
\author{\\
Ali Mostafazadeh\thanks{E-mail address: amostafazadeh@ku.edu.tr}\\
\\ Department of Mathematics, Ko\c{c} University,\\
Rumelifeneri Yolu, 34450 Sariyer,\\
Istanbul, Turkey}
\date{ }
\maketitle
\begin{abstract}
With a view to eliminate an important misconception in some recent
publications, we give a brief review of the notion of a
pseudo-Hermitian operator, outline pseudo-Hermitian quantum
mechanics, and discuss its basic difference with the
indefinite-metric quantum mechanics. In particular, we show that
the answer to the question posed in the title is a definite No.
\end{abstract}

\section{Introduction}

The theory of pseudo-Hermitian operators, as formulated
in\cite{p1,p2-p3,p4-p7,p6}, owes its existence to the author's
efforts to elucidate the origin of the reality of the spectrum of
the $PT$-symmetric non-Hermitian Hamiltonians considered in
\cite{bender1}. It emerged in trying to respond to the question:
``{\em What is the necessary and sufficient conditions for the
reality of the spectrum of a linear operator?},'' and indeed
turned out to achieve its basic goal of understanding the
mathematical structure of the $PT$-symmetric quantum mechanics
\cite{p55}. It also found remarkable applications in other areas
of theoretical physics \cite{cqg,p54,p57,other-ph-app,ahmed}.

Since the initial announcement of the results of \cite{p1} in July
2001, there have appeared dozens of publications exploring various
aspects of the pseudo-Hermitian operators. Among these are a
number of articles \cite{claim,ahmed} reflecting the view that the
notion of a pseudo-Hermitian operator has indeed been known since
as early as the 1940's through the works of Dirac, Pauli, Gupta,
Bleuler, Sudarshan, Lee and Wick \cite{indefinite-ph} and other
authors who developed quantum theories based on a vector space
with an indefinite-metric \cite{indefinite-math}. This view seems
to support the claim that pseudo-Hermitian quantum mechanics can
be reduced to the well-known indefinite-metric quantum theories.
The main purpose of the present article is to show that this is
actually not true. It will be shown that there is a subtle
difference between the notion of a Hermitian operator in an
indefinite-metric vector space (that is admittedly a
`pseudo-Hermitian operator') and the notion of a pseudo-Hermitian
operator as defined in \cite{p1}. This difference which has also
been overlooked in a number of recent publications \cite{claim}
has important conceptual and technical ramifications. These will
also be alluded to here.

\section{Pseudo-, $\eta$-pseudo-, and quasi-Hermitian operators}

We begin our discussion by recalling some basic definitions.
\begin{itemize}\item[]
{\bf Def.~1:} Let ${\cal H}$ be a separable Hilbert space. Then a
linear operator $H:{\cal H}\to{\cal H}$ is said to be {\em
pseudo-Hermitian} \cite{p1} if there exists an invertible,
self-adjoint, linear operator $\eta:{\cal H}\to{\cal H}$
satisfying $H^\dagger=\eta\,H\eta^{-1}$, where $\dagger$ denotes
the adjoint of the corresponding operator.
\end{itemize}
According to Def.~1, the pseudo-Hermiticity of an operator is not
sensitive to the particular form of the operators $\eta$
satisfying $H^\dagger=\eta\,H\eta^{-1}$ but to the existence of
such operators. In fact, it is not difficult to see that either
such an operator $\eta$ does not exist and $H$ is not
pseudo-Hermitian or there are infinitely many $\eta$'s fulfilling
$H^\dagger=\eta\,H\eta^{-1}$ and subsequently $H$ is
pseudo-Hermitian. In the latter case, we shall denote the set of
all such $\eta$'s by ${\cal E}(H)$. For a given diagonalizable
pseudo-Hermitian operator with a discrete spectrum, the problem of
the construction of the most general $\eta\in{\cal E}(H)$ is
addressed in \cite{p4-p7}.

\begin{itemize}
\item[] {\bf Def.~2:} Let ${\cal H}$ be a separable Hilbert space
and $\eta:{\cal H}\to{\cal H}$ be a given invertible,
self-adjoint, linear operator. Then a linear operator $H:{\cal
H}\to{\cal H}$ satisfying $H^\dagger=\eta\,H\eta^{-1}$ is called
{\em $\eta$-pseudo-Hermitian} \cite{p1}.
\end{itemize}
It is essential to observe that, unlike  Def.~1, Def.~2 involves a
fixed operator $\eta$. Clearly, $\eta$-pseudo-Hermitian operators
are pseudo-Hermitian, but not every pseudo-Hermitian operator is
$\eta$-pseudo-Hermitian. This is simply because $\eta$ may happen
not to belong to ${\cal E}(H)$. In summary, the set of
$\eta$-pseudo-Hermitian operators is just a proper subset of the
set of pseudo-Hermitian operators \cite{p8}.

\begin{itemize}
\item[] {\bf Def.~3:} Let ${\cal H}$ be a separable Hilbert space.
Then a pseudo-Hermitian operator $H:{\cal H}\to{\cal H}$ is said
to be {\em quasi-Hermitian} if ${\cal E}(H)$ includes a positive
operator $\eta_+$.
\end{itemize}
Recall that a positive operator is a self-adjoint operator with a
nonnegative spectrum. As the elements of ${\cal E}(H)$ are by
definition invertible, $\eta_+$ is actually a positive-definite
operator, i.e., it has a strictly positive spectrum. Furthermore,
given a quasi-Hermitian operator $H$, the positive operator
$\eta_+$ belonging to ${\cal E}(H)$ is not unique. But any two
such operators $\eta_+$ and $\eta_+'$ are related according to
$\eta_+'=A^\dagger\eta_+ A$ where $A$ is some invertible linear
operator commuting with $H$ \cite{p4-p7}. It is also worth
emphasizing that there are always infinitely many non-positive
elements of ${\cal E}(H)$; a quasi-Hermitian operator is always
$\eta$-pseudo-Hermitian for some indefinite operator $\eta$. Here
by the indefiniteness of $\eta$, we means that neither $\eta$ nor
$-\eta$ are positive-definite operators.

A given pseudo-Hermitian operator may or may not be
quasi-Hermitian. Hence, quasi-Hermitian operators form a proper
subset of the set of pseudo-Hermitian operators. Similarly, the
set of Hermitian operators is a proper subset of the set of
quasi-Hermitian operators:
    $$\mbox{Hermitian}\subset
    \mbox{Quasi-Hermitian}\subset
    \mbox{Pseudo-Hermitian}.$$

The main results of the theory of pseudo-Hermitian operators are
the following spectral characterization theorems.
\begin{itemize}
\item[] {\bf Thm.~1:} Let ${\cal H}$ be a separable Hilbert space
and $H:{\cal H}\to{\cal H}$ be a diagonalizable linear operator
with a discrete spectrum. Then the following are equivalent.
1.a)~$H$ is pseudo-Hermitian; 1.b)~The complex-conjugate of every
eigenvalue of $H$ is also an eigenvalue; 1.c)~$H$ commutes with an
invertible antilinear operator. \item[] {\bf Thm.~2:} Let ${\cal
H}$ be a separable Hilbert space and $H:{\cal H}\to{\cal H}$ be a
linear operator with a discrete spectrum. Then the following are
equivalent. 2.a)~$H$ is quasi-Hermitian; 2.b)~$H$ is Hermitian
with respect to some positive-definite inner product on ${\cal
H}$; 2.c)~$H$ is a diagonalizable operator with a real spectrum;
2.d)~$H$ may be mapped to a Hermitian operator by a similarity
transformation.
\end{itemize}
The proof of Thm.~1 and a slightly stronger variant of Thm.~2 is
given in \cite{p1,p2-p3}. Note also that in view of Thm.~2, Def.~3
is equivalent to the definition of a quasi-Hermitian operator
given in \cite{quasi}.

\section{Pseudo-Hermiticity and the indefinite-metric vector
spaces}

The operators $\eta$ entering the discussion of the
pseudo-Hermitian operators are sometimes called {\em metric
operators}. This is because they may be used to define an inner
product, namely
$\langle\cdot,\cdot\rangle_\eta:=\langle\cdot,\eta\cdot\rangle$,
which is a genuine positive-definite inner product if and only if
$\eta$ is a positive-definite operator. If $H$ happens not to be
quasi-Hermitian, then a positive-definite $\eta$ does not exist,
and the inner product $\langle\cdot,\cdot\rangle_\eta$ with
respect to which $H$ is Hermitian (i.e.,
$\langle\cdot,H\cdot\rangle_\eta=\langle
H\cdot,\cdot\rangle_\eta$) is necessarily {\em indefinite}. This
means that $\langle\cdot,\cdot\rangle_\eta$ satisfies all the
defining relations of an inner product except that there are
$\psi\in{\cal H}$ such that $\parallel\psi\parallel_\eta^2:=
\langle\psi,\psi\rangle_\eta\leq 0$.

Endowing a Hilbert space ${\cal H}$ with a self-adjoint,
invertible, linear operator $\eta:{\cal H}\to{\cal H}$ turns
${\cal H}$ into an {\em indefinite-metric vector space} ${\cal
H}_\eta$. The linear operators that act in this space and are
Hermitian with respect to its indefinite inner product
$\langle\cdot,\cdot\rangle_\eta$ are precisely the
$\eta$-pseudo-Hermitian operators acting in ${\cal H}$. It is the
properties of the $\eta$-pseudo-Hermitian operators for a fixed
$\eta$ that have been studied within the context of the
indefinite-metric vector spaces
\cite{indefinite-ph,indefinite-math}.

The main difference between the approach pursued in the theory of
pseudo-Hermitian operators \cite{p1,p2-p3,p4-p7} over that of the
above-mentioned studies of the indefinite-metric vector spaces
\cite{indefinite-ph,indefinite-math} is that the former does not
involve fixing a metric operator $\eta$ from the outset. This
apparently minor difference has remarkable conceptual as well as
practical implications. The superiority of the former approach
over the latter is reminiscent of that of General Relativity over
Special Relativity. This also reminds one of the following
important lessons of the history of modern physics: 1.~{\em The
geometrical structures underlying physical theories must not be
fixed according to one's wishes or for mere mathematical
convenience}; 2.~{\em Having an arbitrariness in a construction is
an indication of the presence of a symmetry, a quality that is
almost always desirable and often useful.} Ironically, the
tendency to fix a metric operator from the outset and delve in the
intricacies of a fixed indefinite-metric vector space, that has
been the predominant attitude for the past 75 years, is in clear
violation of both these principles.

In contrast to the historical developments leading to the
indefinite-metric quantum theories, the theory of pseudo-Hermitian
operators has been formulated in a way as to incorporate the
freedom in the choice of the metric operator into the basic
structure of the theory. Indeed, the recent application of
pseudo-Hermitian operators in addressing some of the outstanding
open problems of relativistic quantum mechanics \cite{p54,p57} and
quantum cosmology \cite{cqg,p54} makes a direct use of the
arbitrariness of the choice of the metric operator. It is perhaps
the most recent confirmation of the validity of the views
reflected in the above-mentioned lessons of the history of modern
physics.

The particular problems referred to in the preceding paragraph
have to do with the existence and construction of a conserved
positive-definite inner product on the solution space of the
Klein-Gordon equation and its generalizations that arise as the
Wheeler-DeWitt equations in quantum cosmology \cite{qc}. These are
conveniently called Klein-Gordon-type equations \cite{cqg,p54}. It
is well-known that the solution space of such a field equation
admits an invariant indefinite inner product, namely the
Klein-Gordon inner product. The indefiniteness of the latter has
been the fundamental obstacle for devising a probabilistic
interpretation for relativistic quantum mechanics and quantum
cosmology. The theory of pseudo-Hermitian operators provides a
simple and explicit solution for this problem. In the following we
include a brief outline of this solution for the free Klein-Gordon
equation: $ [\partial_t^2+D]\psi(t,\vec x)=0$, where
$D:=-\nabla^2+m^2$, $m$ is the mass, and $c=\hbar=1$.

First, we write the Klein-Gordon equation as a Schr\"odinger
equation for a two-component field. It is well-known \cite{fv}
that the corresponding Hamiltonian $H$ is a
$\sigma_3$-pseudo-Hermitian operator acting in
$L^2(\mathbb{R}^3)\oplus L^2(\mathbb{R}^3)$, i.e.,
$\sigma_3\in{\cal E}(H)$, where $\sigma_3$ is the diagonal Pauli
matrix with diagonal entries $\pm 1$. The Klein-Gordon inner
product corresponds to $\langle\cdot,\cdot\rangle_{\sigma_3}$
which is manifestly indefinite. However, it can be easily shown
that $H$ is diagonalizable and has a real spectrum
\cite{cqg,p54,p57}. Hence according to Thm.~2, ${\cal E}(H)$
includes positive elements $\eta_+$ that can be used to define a
positive-definite inner product
$\langle\cdot,\cdot\rangle_{\eta_+}$. The latter leads to an
explicit expression for a class of positive-definite inner
products on the solution space ${\cal S}$ of Klein-Gordon fields
\cite{cqg}. Because all positive-definite inner products on ${\cal
S}$ are unitarily equivalent \cite{p54}, one may choose any one of
the inner products obtained in this way to develop a probabilistic
quantum theory of first-quantized scalar fields. A particularly
appealing example is the relativistically invariant inner product
\cite{p57}:
    \[(\psi_1,\psi_2):=\frac{1}{2\mu}\int_{\mathbb{R}^3} d^3x~
    \left[\psi_1(\vec x,t)^*D^{1/2}\psi_2(\vec x,t)+
    \partial_t\psi_1(\vec x,t)^*D^{-1/2}\partial_t\psi_2(\vec x,t)
    \right].\]
Endowing ${\cal S}$ with this inner product and completing the
resulting inner product space via Cauchy completion
\cite{reed-simon}, one obtains a separable Hilbert space that
turns out to be most conveniently modelled as
$L^2(\mathbb{R}^3)\oplus L^2(\mathbb{R}^3)$. This allows for an
explicit construction of a novel set of relativistic position
operators and the associated localized and coherent states
\cite{p57}.

It is remarkable that although this problem was formulated in the
late 1920s and examined by some of the founding fathers of both
quantum mechanics (QM) and its extension to indefinite-metric
theories such as Dirac in as early as the 1930s, its solution only
appeared recently. The lack of progress in solving this problem
during the past 75 years may be traced back to the fact that all
the workers on the subject preferred to use the indefinite metric
operator $\sigma_3$ which looked simple and could be related to
the electric charge conservation. It was the recent formulation of
the theory of the pseudo-Hermitian operators (and perhaps the
fortunate ignorance of the author about the early literature on
the subject at the time) that allowed for considering other metric
operators that were unlike $\sigma_3$ positive-definite.

\section{Pseudo-Hermitian and indefinte-metric QM}

Pseudo-Hermitian QM is defined by an auxiliary pseudo-Hermitian
Hamiltonian $H'$ acting in a separable Hilbert space ${\cal H}'$
such that the span ${\cal K}$ of the eigenvectors of $H'$ with
real eigenvalues is nonempty. The physical Hilbert space ${\cal
H}$ and the Hamiltonian $H$ are obtained as follows. First, one
considers the restriction $K$ of $H'$ onto ${\cal K}$ as a linear
operator acting in the complete closure $\bar{\cal K}$ of ${\cal
K}$. Then, by construction $K$ is a densely defined
quasi-Hermitian operator and ${\cal E}(K)$ includes a positive
operator $\eta_+$. Next, one makes a choice for $\eta_+$ (noting
that all choices are unitarily/physically equivalent) and endows
${\cal K}$ with the inner product
$\langle\cdot,\cdot\rangle_{\eta_+}$ so that $K$ may be viewed as
a Hermitian operator acting in this inner product space. ${\cal
H}$ and $H$ are respectively the Cauchy completion of ${\cal K}$
and the closed self-adjoint extension of $K$ to ${\cal H}$,
\cite{reed-simon}.

One can use the auxiliary Hamiltonian $H'$ and the Hilbert space
${\cal H}'$ to formulate an indefinite-metric quantum system. This
is simply done by choosing an arbitrary indefinite $\eta\in{\cal
E}(H')$ and defining the nonphysical Hilbert space ${\cal
H}'_\eta$ to be the indefinite-metric vector space obtained by
endowing ${\cal H}'$ (viewed as a complex vector space) with the
indefinite inner product $\langle\cdot,\cdot\rangle_{\eta}$,
\cite{indefinite-ph}. The physical Hilbert space is then
identified with the subspace of ${\cal H}'_\eta$ that includes
besides the zero vector the elements that have a positive real
norm $\parallel\cdot\parallel_\eta$.

Performing the constructions of the indefinite-metric QM for the
Klein-Gordon equation, one arrives at a physical Hilbert space
that consists of positive-energy solutions. In contrast, the
constructions of the pseudo-Hermitian QM lead to a physical
Hilbert space that includes the positive-, zero-, as well as
negative-energy solutions. This is a concrete evidence that
pseudo-Hermitian QM is not just the well-known indefinite-metric
QM.

{\bf Acknowledgment:} This work has been supported by the Turkish
Academy of Sciences in the framework of the Young Researcher Award
Program (EA-T$\ddot{\rm U}$BA-GEB$\dot{\rm I}$P/2001-1-1).

\end{document}